# A WRF-UCM-SOLWEIG framework of 10m resolution to quantify the intra-day impact of urban features on thermal comfort

Xiaotian Ding[1,2,3,4], Yongling Zhao[4], Yifan Fan[1,2,3*], Jian Ge[1,3], Jan Carmeliet[4]

[1]College of Civil Engineering and Architecture, Zhejiang University, Hangzhou, China
[2]Center for Balance Architecture, Zhejiang University, Hangzhou, China
[3]International Research Center for Green Building and Low-Carbon City, International Campus, Zhejiang University, Haining, China
[4]Department of Mechanical and Process Engineering, ETH Zürich, Zürich, Switzerland

*Correspondence to: Yifan Fan, email: yifanfan@zju.edu.cn

**Abstract**
City-scale outdoor thermal comfort diagnostics are essential for understanding actual heat stress. However, previous research primarily focused on the street scale. Here, we present the WRF-UCM-SOLWEIG framework to achieve fine-grained thermal comfort mapping at the city scale. The background climate condition affecting thermal comfort is simulated by the Weather Research and Forecasting (WRF) model coupled with the urban canopy model (UCM) at a local-scale (500m). The most dominant factor, mean radiant temperature, is simulated using the Solar and Longwave Environmental Irradiance Geometry (SOLWEIG) model at the micro-scale (10m). The Universal Thermal Climate Index (UTCI) is calculated based on the mean radiant temperature and local climate parameters. The influence of different ground surface materials, buildings, and tree canopies is simulated in the SOLWEIG model using integrated urban morphological data. We applied this proposed framework to the city of Guangzhou, China, and investigated the intra-day variation in the impact of urban morphology during a heat wave period. Through statistical analysis, we found that the elevation in UTCI is primarily attributed to the increase in the fraction of impervious surface (ISF) during daytime, with a maximum correlation coefficient of 0.80. **Tree canopy cover** has a persistent cooling effect during the day. Implementing 40% of tree cover can reduce the daytime UTCI by 1.5 to 2.0 ºC. At nighttime, all urban features have a negligible contribution to outdoor thermal comfort. Overall, the established framework provides essential input data and references for studies and urban planners in the practice of urban (micro)climate diagnostics and planning.

# 1. Introduction

Rapid urbanization transforms the urban surface composition, consequently altering its surface energy budget. This transformation leads to urban-rural temperature differences, commonly known as the urban heat island (UHI) effect (Masson et al., 2020; Oke et al., 2017; Zhao, Sen, Susca, Iaria, Kubilay, Gunawarden, et al., 2023). Additionally, global climate change affects cities by altering local air temperature and precipitation patterns, thereby increasing the frequency of heatwave events (Campbell I, Sachar S, Meisel J, 2021; Masson et al., 2020). Collectively, these effects deteriorate outdoor human thermal comfort in urban settings, posing significant risks to human health and well-being. Vulnerable populations such as the elderly and those involved in strenuous physical labor during prolonged exposure (Follos et al., 2021) are particularly at risk. These changes also affect plants and animals within the urban environment (Zipper et al., 2016). Given the increasing urban population, assessing the spatiotemporal variations of urban outdoor thermal comfort and providing implications for urban design has become paramount (Zhao et al., 2023).

City-scale thermal comfort mapping holds significant value for urban planning (L. Chen & Ng, 2011) and assessing health risks during heatwaves (Napoli et al., 2018). However, few studies have achieved such thermal comfort mapping (C. Wang et al., 2020) at this scale. This limitation is due to the challenges associated with evaluating numerous environmental factors, including the surrounding radiant environment, air temperature, relative humidity, wind speed, human physiological characteristics, and thermal resistance of clothing (Potchter et al., 2022). Among these factors, the outdoor radiant environment – resulting from the combined effect of shortwave solar radiation and longwave radiation, significantly influences human energy balance and outdoor thermal comfort during daylight hours (Lindberg et al., 2016). This outdoor radiant environment can be quantified by the mean radiant temperature (MRT), which accounts for all short- and long-wave radiation fluxes, both direct and reflected, impacting the human body (*ASHRAE Fundamentals Handbook 2001 (SI Edition)*, 2001). Despite its significance, assessing MRT with high spatial resolution at the city scale is challenging. It necessitates intricate urban morphological information at a micro-scale, such as urban underlying surfaces, building configurations and orientations, vegetation, and geographic locations. In addition, other factors such as air temperature, relative humidity, and wind speed are influenced by both local urban features and the background climate at larger city- and meso-scale (Ho et al., 2016). Consequently, evaluating the outdoor thermal environment necessitates detailed urban morphological data (including geometry and land cover) at a microscale, along with the consideration of the background meteorological factors.

Past studies (Błazejczyk, 2011; C. Wang et al., 2020) have employed remote sensing data to derive thermal comfort maps at city or larger scales. However, using remote sensing images primarily captures the temperatures of building roofs, neglecting the temperatures of building facades. Moreover, satellite imagery typically has a long revisit cycle (spanning weeks) and a narrow swath (less than 200 km), constraining their utility for city-scale analyses during specific time periods. Global reanalysis data (P. Y. Fan et al., 2022) and mesoscale numerical weather predictions, such as the Weather Research and Forecasting model coupled with the Urban Canopy Model (WRF-UCM), were also used for calculating thermal comfort maps at city-scale (S. Du et al., 2022; X. Wang et al., 2022). While these approaches can provide essential meteorological data for the assessment of thermal comfort, the data often possesses a coarse spatial resolution (greater than 100 m). Moreover, these evaluations typically overlook the influence of detailed urban geometry and fabric, which are crucial for accurately determining a thermal radiant environment and investigating the impact of urban morphology on local thermal comfort.

Revealing the intra-day impact of urban morphology on outdoor thermal comfort can provide valuable implications for urban planning guidelines. While several studies have investigated the intra-day influence of urban morphology at the street scale through field measurements or simulations (Abd Elraouf et al., 2022; Jamei et al., 2016), limited research has focused on the diurnal patterns of urban morphology's influence at a city scale with fine spatial resolution. Zhang et al. (2022) compared the impact of urban morphology on thermal comfort during daytime and nighttime, noting a more pronounced impact during daytime. Wai et al. (2020) investigated the relationship between urban morphology and thermal environment during morning and midday periods, whereas Yu et al. (2020) highlighted the temporal variation in urban morphology's effects on outdoor air temperature throughout a day.

In this study, we propose a methodological framework for evaluating the diurnal pattern of urban microclimate on a city scale, featuring 10-meter spatial resolution and hourly intervals. This enables us to discern the influence of urban morphology in a more robust and statistically significant manner, as corroborated by our findings. This approach couples the mesoscale urban climate simulations with micro-scale outdoor thermal environment simulations. We apply this framework to one of the most densely populated urban areas in China, during a heatwave period. Our objectives are to (a) integrate available urban data needed to simulate mean radiant temperature, and (b) achieve fine-grained city-scale outdoor thermal comfort mapping at hourly intervals spanning multiple diurnal cycles, and (c) evaluate the intra-day variance in the influence of meteorological factors and urban features on outdoor thermal comfort. As a result, our research provides essential input data and references to studies concentrating on urban microclimate diagnostics and modifications.

This paper is organized as follows. In Section 2, we present the methodology for the local-scale urban climate simulation, micro-scale radiant environment simulation, and statistical analysis. The simulation and statistical results are described in Section 3. Section 4 provides a comprehensive discussion of these results. Finally, conclusions are drawn in Section 5.

## 2. Methodology
*2.1. Framework for city-scale thermal comfort mapping*

We first employ the **mesoscale urban climate simulation**, WRF-UCM, to derive background climate data, as shown in Figure 1. The **micro-climate model**, the Solar Long Wave Environmental Irradiance Geometry model (SOLWEIG) (Lindberg et al., 2008a), is then utilized to simulate the most important factor affecting human thermal comfort: the mean radiant temperature (MRT) at hourly intervals. The impact of urban morphology on MRT is considered using **integrated urban data**. Subsequently, the thermal comfort index, i.e., the Universal Thermal Climate Index (UTCI) (Bröde et al., 2009), is calculated to characterize the overall human sensation to the micro-scale thermal environment. This calculation is based on a $6^{th}$ order polynomial approximation of the original algorithm (http://www.utci.org/), represented by the function $f$ (AT, RH, WS, MRT), where the inputs are air temperature (AT), relative humidity (RH), wind speed (WS), and mean radiant temperature (MRT). Variations in background climate conditions (AT, RH, and WS) within each WRF simulation grid exert a limited impact on thermal sensation compared to MRT. Consequently, we determine UTCI at the same spatial resolution as the micro-scale MRT instead of aligning it with local-scale climate data. Additionally, we evaluate the dependence and contribution of urban morphological parameters on UTCI through correlation and regression analysis, offering valuable insights for future thermal-environmentally guided urban planning.

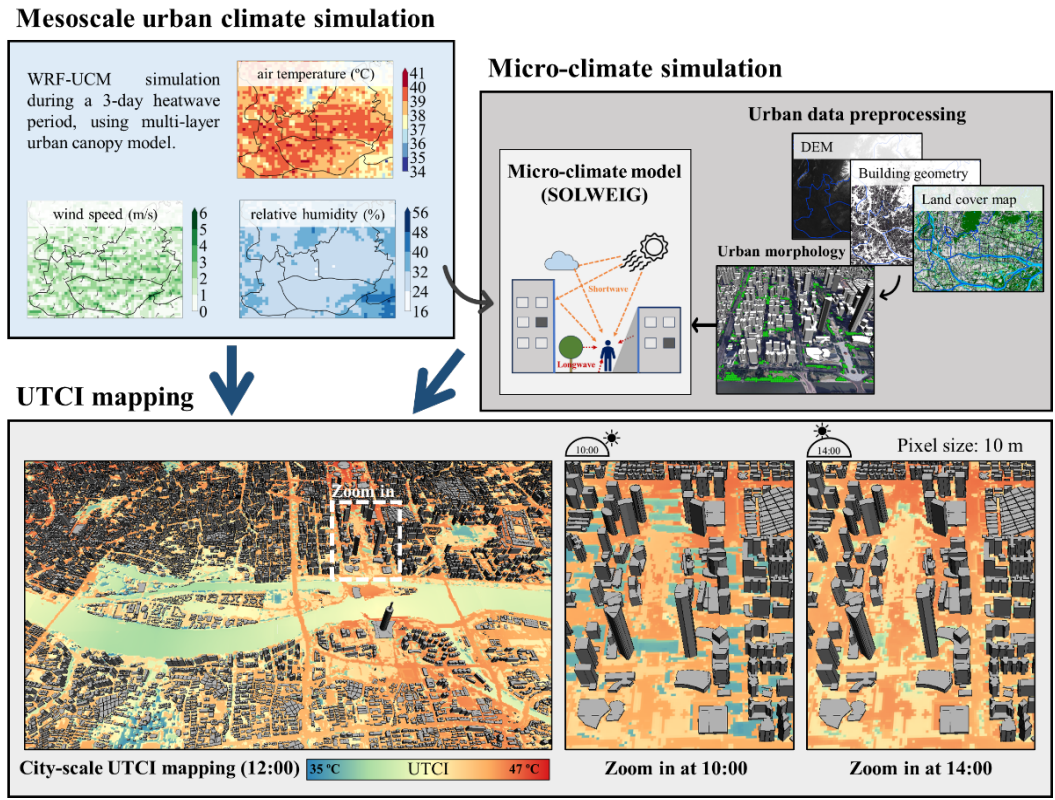

**Figure 1.** Schematic representation of the framework for city-scale thermal comfort mapping and analysis. Effect of building shade on local UTCI can be observed in the zoom-in at 10:00 and 14:00.

*2.2. Study area and local-scale urban climate simulation*

We apply this framework to the densely populated urban areas of Guangzhou (23.1 N, 113.3 E), China, as shown in Figure 2a. This area encompasses four administrative districts, covering a total area of 279.6 square kilometers. The majority of Guangzhou's population, approximately 6.4 million people (or 23,000 people per square kilometer), resides within this area. The simulation utilized the WRF model version 4.4, integrated with the building effect parameterization (BEP) and building energy model (BEM) scheme (Martilli et al., 2002; Salamanca et al., 2010). Four nested domains were used in this study, as shown in Figure 2a, and the resolution of each domain is 13.5 km, 4.5 km, 1.5 km, and 0.5 km for each domain. The inner domain (Domain04) consists of a 121 × 121 grid matrix with a grid cell size of 500 m, while the study area occupies a 54 × 44 grid matrix within this inner domain (Table B1). The simulation was validated using data from 72 municipal automatic weather stations, as presented in Figure 2b. Comprehensive details on the model configuration are available in Appendix A, with the evaluation process detailed in Appendix B.

Guangzhou is characterized by a subtropical climate with warm and humid summers. Based on climatological data from 1991 to 2020, the hottest months are July and August, during which the average daily air temperatures is 28.8 °C and the average daily maximum reaches 33.3 °C. This study focuses on a three-day heatwave, defined as a span of more than three consecutive days where daily maximum temperatures surpass 35 ºC (Sun et al., 2018). The selected period of analysis runs from 8 a.m. on August 20 to 8 a.m. on August 23, 2021, representing the hottest days in the past three years. During this period, sunrise in Guangzhou is at 06:50, and sunset is at 18:50. Observation data reveals that the mean daily air temperature during this heatwave exceeds 30.6 ºC, with the daily maximum soaring above 35.9 ºC. Both these daily mean and maximum readings are notably higher than the climatological (1991-2020) averages, by 2 ºC and 2.7 ºC respectively. The conditions during this time also feature a clear sky and calm synoptic wind, with an average wind speed of 1.4 m/s (standard deviation of 0.2 m/s).

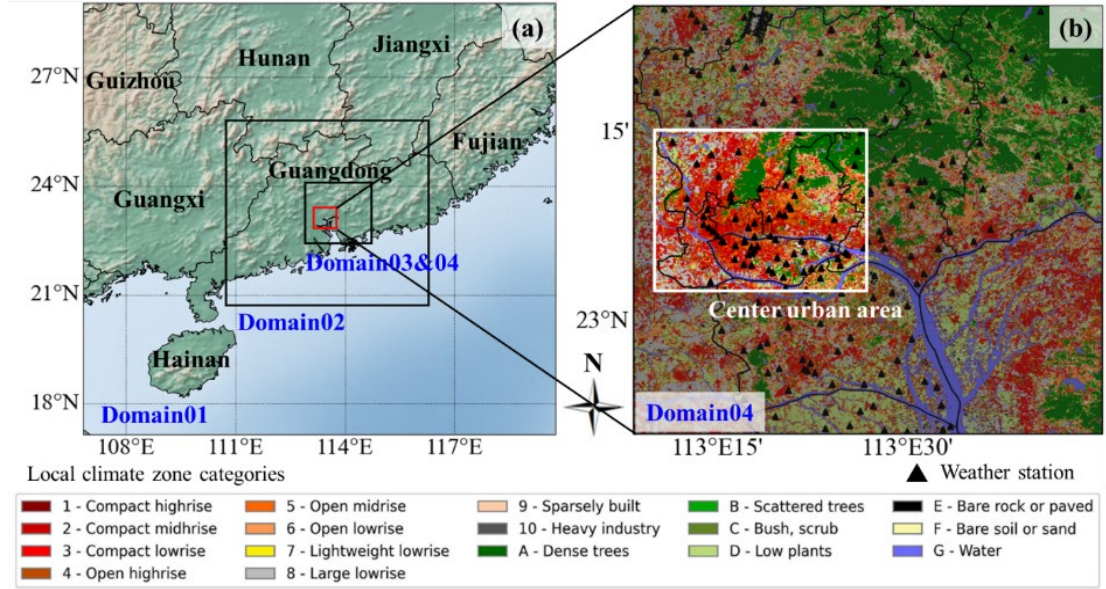

**Figure 2**. Location of the study area. (a) Nested domains (01-04) for WRF simulation. (b) The local climate zone (LCZ) map with the study area highlighted in the white frame. The location of Guangzhou AWS is marked with black solid triangles.

*2.3. local-scale radiant environment simulation*
*2.3.1 Urban morphological data acquisition*

In this study, we conducted the simulations at a 10-meter resolution to align with the available urban data. We acquired a high-resolution land cover map, a digital elevation model (DEM), and three-dimensional (3D) building information from open-source data. Detailed description regarding these urban morphological data is presented in Table 1. The DEM for Guangzhou was obtained from the ALOS PALSAR Radiometrically Terrain-Corrected (RTC) product (Takaku et al., 2014). For the collection of three-dimensional building information data, we utilized web-mapping platforms, such as OpenStreetMap (OSM, https://www.openstreetmap.org/) and Baidu digital maps (https://map.baidu.com/). The original building data includes footprints and the number of floors for each building. In this study, we assume an average floor height of 3 m. All data were processed using the pre-processor tool available in the Urban Multi-scale Environmental Predictor (UMEP) plugin in the QGIS platform [QGIS Geographic Information System, QGIS Association, http://www.qgis.org]. This processing yielded intermediate parameters such as the sky view factor (SVF), wall height and aspect, and the respective shadow patterns for each hour of the day.

In order to consider the effects of land surface cover, we employed a land cover map derived using a deep-learning-based image classification method applied to Google Earth imagery (Ding et al., 2022; Y. Fan et al., 2021). The Google Earth imagery was classified into eight land surface categories: "Building", "Road", "Other impervious surface", "Tree", "Low vegetation", "Bare lands", "Water" and "Shadow" for shaded urban surfaces, as illustrated in Figures 3a and 3b. The classification's overall accuracy for the Guangzhou area, defined as the ratio of correctly predicted pixels to the total pixels, stands at 84.4% on test dataset (Ding et al., 2023). This accuracy was deemed accepted for our study, aligning with the standards set by the National Land Cover Database (NLCD) (Homer et al., 2020). In the land cover map, the "Tree" category provides a two-dimensional representation of the tree canopy, capturing details about its location and spread. From this, we designated a virtual tree at intervals of 10 m on the derived "Tree" cover map, resulting in an idealized 3D tree canopy map, as shown in Figure. 3c. Each of these virtual trees is uniformly given the same geometric shape, with a trunk height of 3 m, a total tree height of 8 m, and a canopy radius of 5 m. For compatibility with the calculation grids, the canopies of these virtual trees were shaped as squares rather than circles.

**Table 1**

Original urban morphological data used in this study.

| Variables | Description | Spatial resolution | Data source |
|---|---|---|---|
| DEM | Digital elevation model. | 12.5 m | ALOS PALSAR RTC (12.5 m) product |
| 3D Building information | 3-dimensional building information (polygons): the footprint and the number of floors of each building. | - | OpenStreetMap and Baidu map |
| Land cover map | Urban land cover map containing eight land cover categories which are derived from Google Earth image using a deep learning-based method. | 2 m | Derived from Google Earth image |
| 3D Tree canopy map | The location, tree trunk height, and canopy size information. | - | Generated based on the "Tree" category in the land cover map |

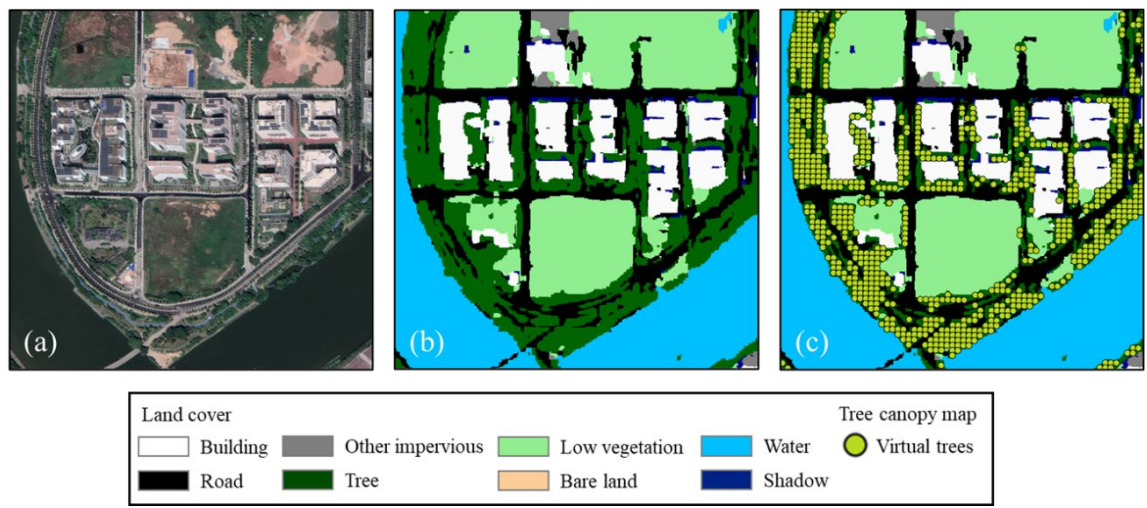

**Figure 3.** Demonstration of planting virtual trees based on derived land cover. (a) clip of the Google imagery of Guangzhou during the year of 2019-2021. (b) Land cover map derived from Google Imagery. (c) The location of virtual trees.

*2.3.2. Simulation of mean radiant temperature*

The mean radiant temperature (MRT) within the study area was simulated on an hourly basis using the Solar Long Wave Environmental Irradiance Geometry model (SOLWEIG) (Lindberg et al., 2008) implemented in the UMEP. A detailed description of the SOLWEIG model can be found in Appendix C. The SOLWEIG model takes two categories of input data: urban spatial information and meteorological data. The urban spatial data can be further divided into two sub-categories: urban surface models, which comprise height information, and ground cover data, as listed in Table 2. All spatial data used during the calculation were resampled into the same spatial resolution of 10 m. The urban surface models incorporate the digital elevation model (DEM), along with building and vegetation DSM. The land cover map, derived from Google Earth imagery, was further allocated with specific surface properties, as listed in Table C1. It is noteworthy that areas beneath trees are presumed to have grass as the ground cover (Lindberg et al., 2018). The required meteorological data are direct, diffuse and global shortwave radiation, air temperature, and relative humidity. For this study, these data were extracted from the WRF simulation.

**Table 2**
Description of input data for SOLWEIG.

| Data category | Description | Spatial resolution | Data source |
| --- | --- | --- | --- |
| Meteorological data | Including hourly direct, diffuse and global shortwave radiation, air temperature, and relative humidity | 500 m | WRF simulation |
| Digital surface models | Ground, building and tree height information | 10 m | DEM, vegetation and building DSM |
| Ground cover | Ground cover including grass, building, impervious dark asphalt, water and bare soil | 10 m | Land cover map |

*2.5. Evaluating the influence of urban feature*

*2.5.1. Calculation of urban features*

To investigate the influence of urban morphology on local thermal environment, this stud employed Pearson's correlation and regression analysis. All grid cells in the WRF simulation within the study area, totalling 2376 grids in a 54 × 44 grid matrix, were regarded as the buffer zone grids (500 m × 500 m), or data points for the statistical analysis. The chosen scale aligns with recommendations from prior studies (Su et al., 2022; Yang et al., 2020) when examining the impact of building-up areas. As our study focuses specifically on urban built-up areas, 344 data points in the mountainous region (average elevation > 50 m) were excluded. This exclusion resulted in a final dataset comprised 2032 data points (2376 – 344 = 2032) for subsequent analysis. During the analysis, the mean UTCI value of each buffer zone grid served as the dependent variable, while meteorological and urban morphological parameters within the same grid were considered as independent features. The description and respective calculation equations for these variables are illustrated in Figure 4. Additionally, relevant statistical data is presented in Table 3.

As depicted in Figure 4a, the sky view factor (SVF) was derived based on the vegetation and building digital surface model (DSM), which combines the DEM, 3D building and tree canopy, using the UMEP pre-processor tool. The floor area ratio (FAR) and building density (BD) and were calculated using the building footprints and the building height information obtained from the 3D building data, as shown in Figure 4b, c. Using the Google Earth derived land cover, we calculated five two-dimensional (2D) parameters, as shown in Figure 4d-f, including the fraction of impervious surface (ISF), road surface (RSF), green surface (GSF), water surface (WSF) and tree canopy cover (TCF). The methodology for calculating SVF is documented in detail in Lindberg et al. (2010). All these calculations were performed on the QGIS platform.

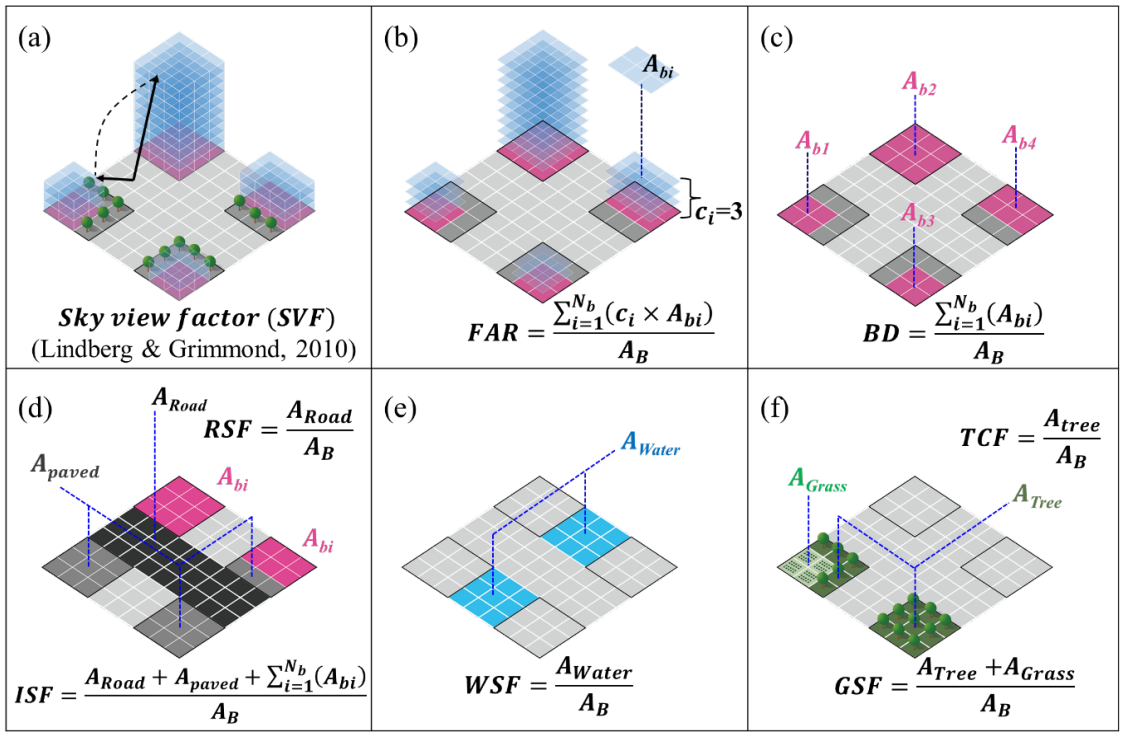

**Figure 4.** Illustration of the selected urban morphological variables: (a) the sky view factor or 'SVF' based on the method (Lindberg & Grimmond, 2010). (b) floor area ratio or 'FAR'. (c) building density or 'BD'. (d) road surface fraction and impervious surface fraction. (e) water surface fraction or 'WSF'. (f) the tree canopy fraction or 'TCF' and green surface fraction 'GSF'.

**Table 3**

Descriptive statistics of the selected urban morphological variables.

| Variables | Mean | Standard deviation | 5th percentile | 95th percentile | Min | Max |
|---|---|---|---|---|---|---|
| *ISF* | 58.5 | 25.6 | 11.0 | 92.2 | 0.0.  | 99.9 |
| *RSF* | 17.1 | 10.0 | 2.9 | 35.3 | 00 | 72.0 |
| *GSF* | 31.9 | 20.4 | 6.2 | 72.9 | 0.1 | 100.0 |
| *WSF* | 9.0 | 18.3 | 0.0 | 56.1 | 0.0 | 99.5 |
| *TCF* | 18.2 | 18.0 | 1.3 | 56.6 | 0.0 | 99.9 |
| *BD* | 19.9 | 14.9 | 2.7 | 45.8 | 0.0 | 67.1 |
| *FAR* | 2.9 | 2.8 | 0.0 | 8.4 | 0.0 | 15.8 |
| *SVF* | 0.81 | 0.15 | 0.51 | 0.97 | 0.05 | 0.99 |

*2.5.2. Correlation and regression analysis*

Following the analysis approach by Cao et al. (2021), the Pearson's correlation coefficients (R) was used to measure the linear correlation between dependent variable (UTCI) and independent variable (meteorological and morphological parameters) over the 3-day simulation period. The correlation coefficients for each hour were calculated by dividing the covariance of the two variables by the product of their standard deviations. The significance of the coefficient was assessed using p-values (Witte & Witte, 2017), a significance threshold of 0.05 was applied. If the p-value was less than 0.05, the correlation between the two variables was considered statistically significant. Then, the variance inflation factor (VIF) was used to examine collinearity and select variables for the regression analysis.

The Random Forest (RF) regression was adopted to assess the contribution of urban features to outdoor thermal environment in this study. Urban features with VIFs < 10 are used for the regression analysis. Besides, considering the influence of the local meteorological condition on thermal comfort, we also include meteorological parameters as the input features. As a result, two meteorological parameters (AT and WS) and five urban morphological parameters (BD, FAR, RSF, TCF, WSF) with no multicollinearity were selected. Although the WSF also has no collinearity with those selected parameters, it is not involved in the regression analysis because the thermal comfort analysis over water surface does not have significant practical implications. In order to avoid overfitting, data points were randomly separated into train and test sets in an 8:2 ratio.

To measure the relative contribution of each input variable, we adopted the Shapley Additive Explanations (SHAP) value, which is a widely used approach (Gao et al., 2023;

Han et al., 2022; Lundberg & Lee, 2017; Wu et al., 2022) to explain the tree-based model prediction by calculating the relative contribution of each feature to the prediction. Locally, a positive or negative SHAP value (≠ 0) indicates that the feature point has a positive or negative impact on the prediction values. Globally, the average absolute SHAP or |SHAP| value for each feature represents its relative importance and features with larger |SHAP| values are more important. Besides, the partial dependence plot (PDP) was applied to further interpret the model by visualizing how a machine learning model's predictions change as a specific feature is varied, while keeping all other features constant. This helps visualize the feature's impact on the model's output (Gao et al., 2023; Han et al., 2022). PDP can intuitively reveal the linear or nonlinear effect each feature has on the predicted outcome of a regression model, assuming other characteristics remain unchanged. A flat PDP indicates that the feature is not important, and the steeper the PDP, the greater the contribution the feature has on the output.

**3. Results**

*3.1. Spatial and temporal characteristics of UTCI*

Based on the validated mesoscale WRF simulation and micro-scale SOLWEIG simulation, the hourly UTCI maps were generated during the heatwave period, as shown in Figure 5. Furthermore, we obtained Global Human Settlement (GHS) population data in a 100 m grid for the year 2020(Maffenini et al., 2023), as shown in Figure 5d. The MRT and UTCI values at each urban grid (mean elevation < 50 m) were spatially averaged for each hour to generate a diurnal variation, as illustrated in Figure 6a. By overlaying the population grid data with the UTCI maps, we computed the percentage of the population exposed to heat stress. The statistical descriptions of UTCI and the population exposed to heat stress are illustrated in Figures 6b and 6c, and detailed in Table 4.

The spatial variation in daytime UTCI among different land cover types is more pronounced than at night. These findings are consistent with previous research (Kong et al., 2022; C. Wang et al., 2020). The spatial distribution of daytime UTCI generally reflects the distinct characteristics of various land cover types, as shown in Figure 5b, e. Impervious road surfaces are typically hotter than areas shaded by trees or buildings, especially during the day. However, after sunset, the spatial variation in UTCI across the study area and its association with land cover types diminish. This change is attributed to the absence of incoming shortwave radiation and a notable drop in nighttime surface temperatures over urban areas, evident in the colored area in Figure 5c, f. Both UTCI and MRT exhibit a distinct diurnal variation, characterized by steep rises and falls during the morning hours (07:00 - 10:00) and in the evening (16:00 - 19:00), as depicted in Figure 6a. This pattern suggests swift changes in the radiant environment shortly after sunrise and before sunset.

To investigate the intra-day pattern of heat exposure, we divided one day into four time periods: morning period (07:00 - 10:00), noon period (10:00 - 16:00), evening period (16:00 - 19:00), and nighttime (19:00 - 07:00), as listed in Table 4. During the noon period, the average UTCI is 42.6 ºC (with a standard deviation of 0.2 ºC), surpassing the threshold (38 ºC) that distinguishes strong heat stress from very strong heat stress. As illustrated in Figure 6b, nearly all residents (99.2%) are exposed to very strong outdoor heat stress during this time. Even during the nighttime period, more than 96% of the population endures strong heat stress (32 ºC <UTCI<38 ºC), as shown in Figure 6c. Compared with air temperature, thermal comfort indexes, such as UTCI, correlate more closely with heat-induced mortality (Napoli et al., 2018). The results indicate that the human-perceived temperature, which represents the equivalent temperature causing the same dynamic physiological response, is on average 4.8 ºC higher than the air temperature (Figure B1). Therefore, this approach can effectively capture the temporal and spatial variations of thermal comfort in urban areas. The findings can be used for evaluating city-scale heat exposure, making the approach valuable for evaluating heat-related risks and forecasting human thermal stress in real-time settings (Napoli et al., 2021).

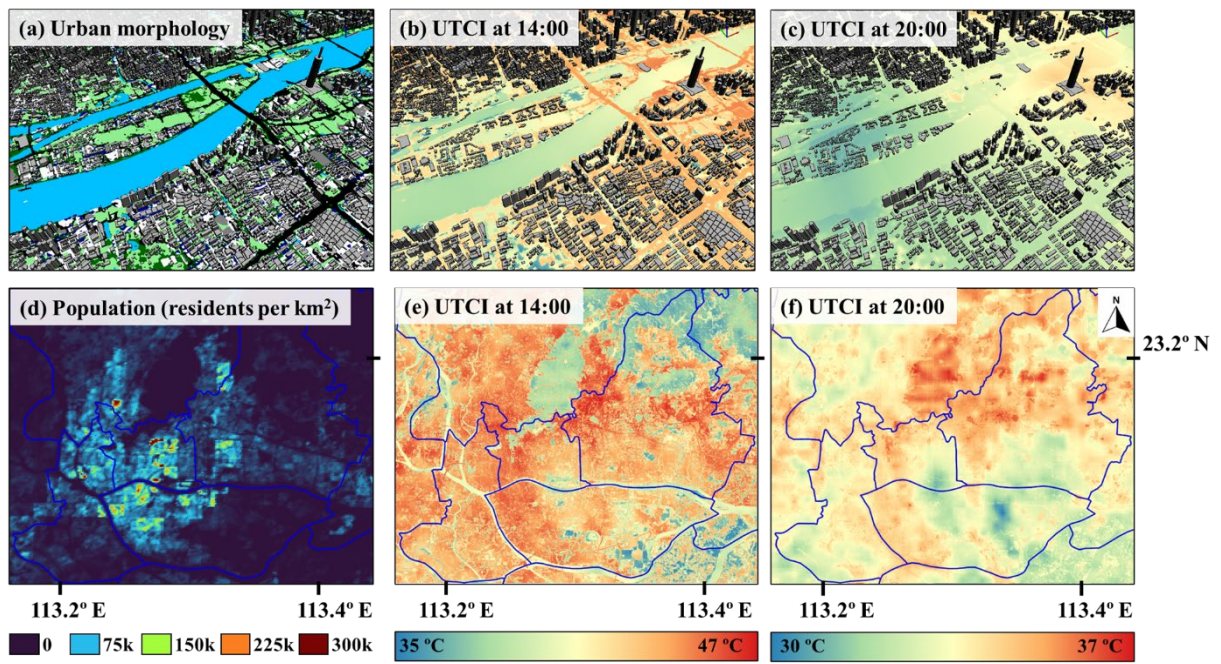

**Figure 5.** Spatial distribution of UTCI. (a, b, c) The comparison of urban morphology and UTCI map at 14:00 and 20:00. (d) Population distribution. (e, f) The UTCI map of the study area, at 14:00 and 20:00.

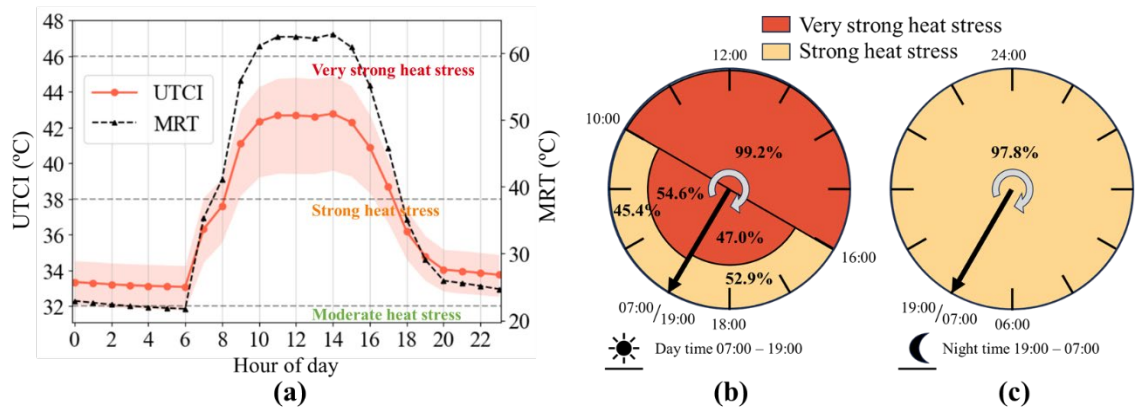

**Figure 6.** (a) Temporal variation of mean MRT and UTCI in urban grids (mean elevation < 50m). The colored area denotes 95% confidence intervals of UTCI. Percentage of population expose to heat stress at each time period, including daytime period (b) from 07:00 AM to 19:00 PM and nighttime period (c) from 19:00 PM to 07:00 AM.

**Table 4**
The statistical description of UTCI and population expose to heat stress at each time period. The standard deviation (Std.) is calculated to assess the temporal variation of UTCI during each period. (Guangzhou's sunrise and sunset occur at 06:50 and 18:50 during the heat wave.)

| Name | Time period | Mean UTCI (Std.) (ºC) | Percentage of population under extreme heat stress (38 ºC < UTCI < 46 ºC) | Percentage of population under very strong heat stress (38 ºC < UTCI < 46 ºC) | Percentage of population under strong heat stress (32 ºC < UTCI < 38 ºC) | Percentage of population under moderate heat stress (32 ºC < UTCI < 38 ºC) |
|---|---|---|---|---|---|---|
| Morning period | 07:00-10:00 | 37.0 (3.3) | 0.00 % | 54.62 % | 45.35 % | 0.03 % |
| Noon period | 10:00-16:00 | 42.6 (0.2) | 0.45 % | 99.22 % | 0.33 % | 0.00 % |
| Evening period | 16:00-19:00 | 37.6 (2.7) | 0.00 % | 47.07 % | 52.90 % | 0.03 % |
| Nighttime | 19:00-07:00 | 33.6 (0.5) | 0.00% | 0.00 % | 96.77 % | 3.23 % |

*3.3. Correlation between meteorological/morphological parameters and UTCI*

The correlation coefficients and associated p-values were averaged for every hour of the day and are illustrated in Figure D1 of Appendix D. A distinct diurnal pattern emerges from the results, as highlighted in Figure 7. For meteorological parameters, as shown in Figure 7a, both MRT and AT are positively correlated with UTCI. This suggests that increased MRT and AT values correspond with elevated UTCI values. In contrast, wind speed and relative humidity both correlate negatively with UTCI. During daytime (06:00-19:00), the most robust correlation is between MRT and UTCI, with an average coefficient of 0.79. This underlines that UTCI is predominantly influenced by MRT during this period. Following MRT, AT, RH and WS also demonstrate correlations with UTCI, but these associations are less strong. On the contrary, during nighttime hours (19:00-07:00), the correlation pattern changes. In the absence of sunlight, air temperature becomes the primary correlating factor with UTCI, with an average coefficient value of 0.55, emphasizing its augmented role in influencing UTCI values at night. Both wind speed and relative humidity exhibit much less dominant but considerable correlations with UTCI throughout this period.

Regarding urban morphological parameters, the correlations with UTCI are generally more pronounced during the daytime thank nighttime, as shown in Figure 7b-e. During the early morning hours (07:00 and 08:00), areas with a larger proportion of buildings (indicated by higher values of ISF, BD, and FAR) and a smaller SVF tend to have a better thermal environment. This suggests that both high-rise buildings and tree canopies can create a cooling effect on the outdoor thermal environment in the early morning. After 9:00, UTCI shows positive correlations with factors related to the man-made environment, including ISF, BD, RSF and FAR (Figure 7b, c). On the other hand, UTCI exhibits negative correlations to natural surface proportions, such as TCF, GSF, and WSF (Figure 7b, d). This indicates that, during that period, densely built-up areas with a higher proportion of buildings and roads tend to have poorer thermal comfort compared to areas with a higher proportion of natural surfaces, particularly areas with more trees. Open areas with a larger SVF still tend to have higher UTCI values due to the absence of shade effects. During nighttime (19:00 - 07:00), UTCI only demonstrates weak correlations with 3D morphological parameters, including SVF, FAR, and TCF (Figure 7b). Other 2D parameters almost exhibit no correlation with UTCI, except for WSF, which shows a weak negative correlation. Water surfaces have a negative correlation with UTCI, but it is weaker than during daytime. The open areas with higher SVF now tend to have better thermal comfort, due to improved ventilation and greater radiant heat transfer with the cooler nighttime sky.

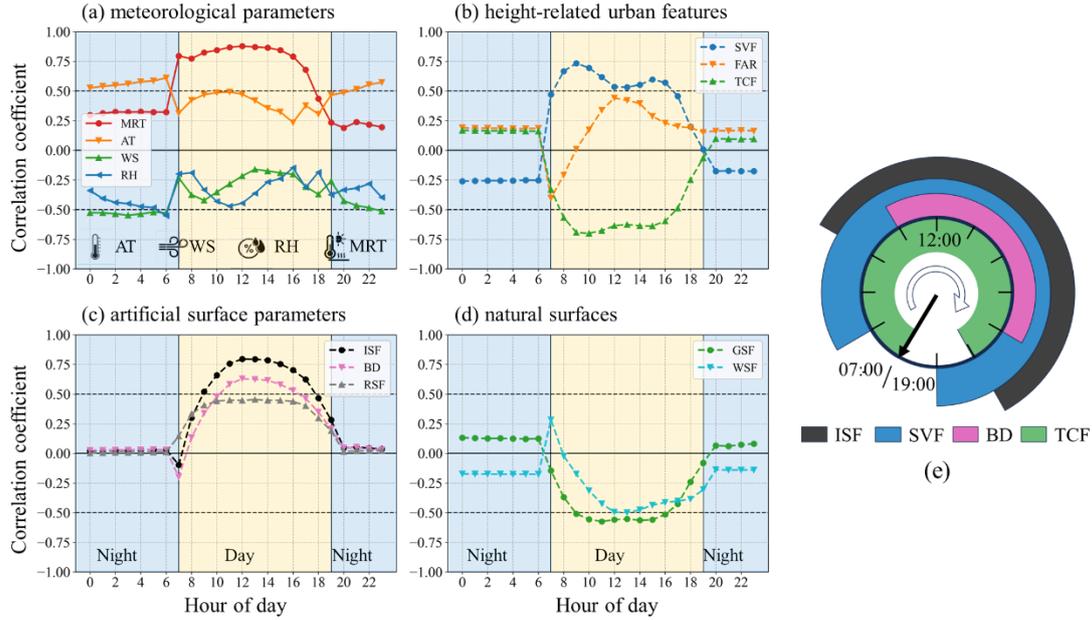

**Figure 7.** Diurnal patterns of mean Pearson's correlation coefficients between thermal comfort and (a) meteorological parameters, i.e., MRT, AT, WS and RH, (b) height-related urban features, i.e., SVF, FAR and TCF, (c) artificial surface parameters, i.e., ISF, BD and RSF, and (d) natural surfaces, i.e., GSF and WSF. (e) The most significant morphological impact factors (r ≥ 0.5) during daytime from 7:00 AM to 19:00 PM.

*3.4. Quantify the contribution of urban morphological parameters*

During the regression analysis, random forest (RF) regression models for UTCI are fitted for each hour using two meteorological parameters (AT and WS) and five urban morphological parameters (BD, FAR, RSF, TCF). The contribution of each input variable to the spatial heterogeneity of UTCI is measured by the |SHAP| values. The |SHAP| values and the regression scores are averaged at each hour, as shown in Figure D2. The diurnal pattern of the |SHAP| values is depicted in Figure 8. For meteorological parameters, the result shows air temperature has a larger contribution to UTCI than wind speed in most hours, as shown in Figure 8a. This is probably because the wind speed is typically low (Figure B1c) during this heatwave event, making air temperature a more influential factor. As for urban morphological parameters, their contribution shows a distinct daytime variation and similar nighttime impact. During the daytime, the TCF has the largest average contribution to UTCI, with a mean |SHAP| value of 0.56. As shown in Figure 8b, the impact of TCF gradually increases after sunrise, reaching its first peak between 9:00 to 11:00 AM. This is followed by a sharp decline, hitting a trough at 13:00 PM. Subsequently, the impact of TCF rises again, reaching a second lower peak at 14:00 and 15:00 PM, and diminishes afterwards. The second most impactful factor is BD, with a mean |SHAP| value of 0.26. The impact of BD ascends after 11:00 AM, and descends after reaching its peak at 13:00 PM. The diurnal pattern of RSF closely resembles that of BD, but with a significantly lesser influence. The impact of FAR is only evident during the early morning hours, peaking at 7:00 AM and gradually diminishing over the next two hours.

Additionally, partial dependence plots (PDPs) are employed to further quantify the impact of urban features on the RF regression model outputs. PDPs of the three most influential urban features (TCF, BD, and FAR) at four specific hours (09:00, 13:00, 15:00 and 20:00) are drawn (Figure 9). As shown in Figure 9a, g, increasing the TCF from 0 % to 40 % can reduce UTCI by more than 2.0 ºC at both hours. The impact diminishes when the solar angle is near its highest (13:00) or when the solar radiation flux is at its lowest (7:00 and 18:00). However, even at 13:00, the urban tree canopies still have a strong cooling effect, with a mean |SHAP| value of 0.47, reducing UTCI by approximately 1.8 ºC when TCF is over 40% (Figure 9d). It indicates that increasing urban tree canopy cover is an effective strategy to improve daytime outdoor thermal comfort, more so than modifying other urban features.

The contributions of BD and RSF follow a similar diurnal pattern, with both showing strong contributions at 13:00 and a greater impact in the afternoon (12:00 - 18:00) than in the morning (7:00 - 11:00). However, the magnitude of BD's impact surpasses that of RSF, with a daytime mean |SHAP| value of 0.22. This result can be attributed to two factors: firstly, the surface temperature of building walls and roads are nearing their maximum deviation from the air temperature, and secondly, the solar angle is high around noon, minimizing the shade effects of building and tree canopies. Specifically, as illustrated in Figure 9e, the partial dependence plot of BD illustrates a non-linear relationship with UTCI. An increase from 0 % to 30 % has the most pronounced positive effect on UTCI, raising it by about 2.4 - 2.6 ºC. However, further increments in BD beyond 30 % have a negligible impact on UTCI. On the other hand, the shading effect of buildings can contribute to reducing UTCI in the early morning, with FAR having a mean |SHAP| value of 0.31 at 7:00, as illustrated in Figure 9c. Taller buildings with larger FAR values contribute more significantly to reducing UTCI. This effect, however, diminishes quickly in three subsequential hours, as shown in Figure 9(f, i), indicating the cooling effect of high-rise buildings is only significant in the early morning when the surface temperature of building walls and roads remain low. These results suggest that urban areas with high-rise building having larger FAR values but lower BD may offer better thermal environments than areas dominated by low-rise buildings, which would exhibit higher building density (BD).

During nighttime, even with relatively smaller mean |SHAP| values than during the daytime, TCF remains the most influential morphological feature. Urban trees could have a nighttime warming effect on thermal comfort, which is consistent with previous findings (T. Chen et al., 2021; Eniolu et al., 2017). While tree canopies enhance daytime thermal comfort through shading, they contribute to nighttime warming via radiative trapping. Nevertheless, this adverse effect is minimal compared to their daytime cooling

effect. As shown in Figure 9j, even with TCF reaching 60 % in the buffer zone, the UTCI only rises by about 0.2 ºC to 0.3 ºC at 20:00. FAR also has a slight contribution to nighttime warming, as shown in Figure 9l, but its magnitude is considerably smaller compared to TCF. The impact of other features during nighttime is negligible, with their |SHAP| values lower than 0.05. The variance in BD barely affects nighttime UTCI, as shown in Figure 9k.

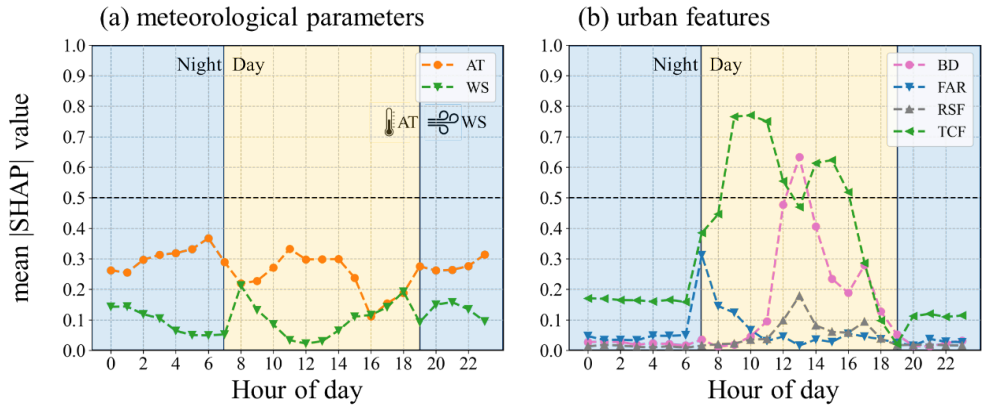

**Figure 8.** The diurnal pattern of mean |SHAP| values for each input feature: (a) meteorological features, i.e., the air temperature (AT) and wind speed (WS), and (b) the urban features, i.e., BD, FAR, RSF, and TCF.

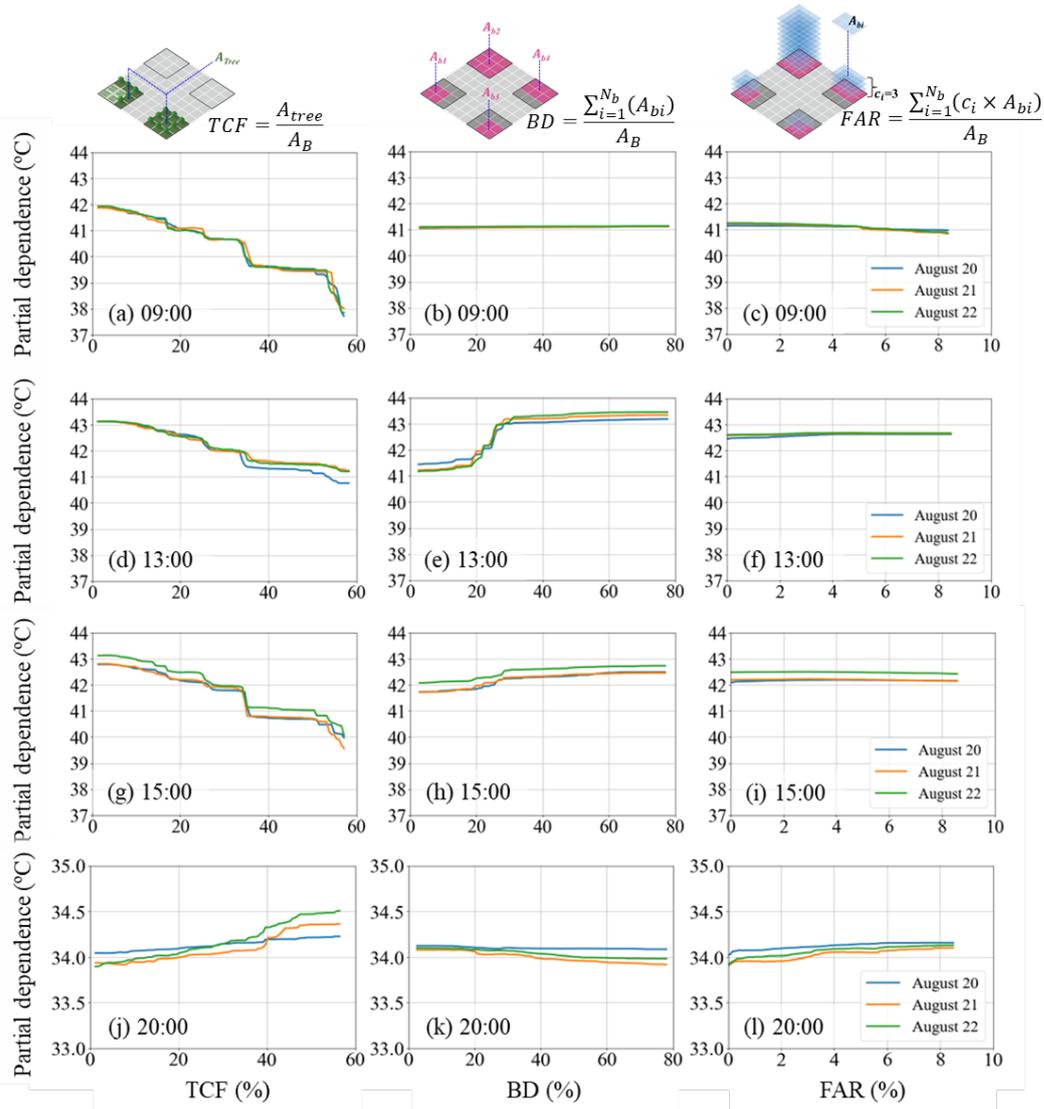

**Figure 9.** Partial dependence plot of input features (TCF, BD, and FAR) at 09:00 (a-c), 13:00 (d-f), 15:00 (g-i) and 20:00 (j-l) on different dates (from 20[th] August to 22[nd] August).

## 4. Discussion

The evaluation of urban microclimate is crucial for urban planning and formulating urban heat mitigation strategies. Especially during heatwaves, the excessive heat stress in urban area becomes a severe threat to both human and plant health. Therefore, this study focusing on achieving high-resolution thermal comfort mapping at city-scale and elucidating the roles of the urban morphological feature on nighttime and daytime urban thermal environment at city-scale during a heatwave.

To obtain micro-scale thermal environment at high spatial and temporal resolutions, fine-grained urban morphology data，including land cover, urban geometry, and tree canopies，were integrated and employed in our established WRF-UCM-SOLWEIG simulations. Considering the limited impact of variations in background climate conditions within each WRF simulation grid, UTCI was calculated at the same spatial resolution as the micro-scale MRT. Therefore, the city-scale thermal environment diagnostics can be achieved at a 10 m spatial resolution. The approach is highly applicable for assessing heat exposure in urban areas, facilitating the evaluation of heat-related risks and the forecasting of human thermal stress in real-world setting. It is essential to note that we define the spatial resolution of our final thermal comfort maps to be in line with the simulated mean radiant temperature (MRT) instead of local-scale meteorological parameters. This decision was based on the premise that MRT primarily governs human thermal sensation. However, our findings suggest that this approach might cause a slight underestimation in UTCI spatial variations. This could be particularly noticeable during nighttime when background climate factors exert a more significant influence, as indicated by our analysis results.

In this study, our primary focus is to demonstrate the proposed framework for city-scale thermal comfort mapping. However, it is worth noting that improvements on spatial resolution are feasible when finer-grained urban geometry data is available, allowing for resolution as fine as 1 m or better (Kántor et al., 2018; Lindberg et al., 2008b). In terms of enhancing future city-scale thermal comfort mapping, acquiring meteorological data at a higher spatial resolution through innovative methods, such as statistical approaches aided by machine learning models (Ding et al., 2023), presents a more promising and cost-effective approach compared to integrating WRF with CFD simulations. Additionally, while the evaluation of the SOLWEIG shows good agreement between simulated and measured values (Lindberg et al., 2008a, 2016), the simplified nocturnal surface temperature parameterization in the model may slightly underestimate the impact of land surfaces during the night. More advanced surface temperature parameterization schemes are needed to improve accuracy.

The case study conducted in Guangzhou, China, sheds light on the profound relationship between urban morphological and thermal comfort. During daytime, the contribution of morphological parameters on thermal comfort is significant. Among those analyzed urban features, tree canopies emerge as the most influential factor in enhancing daytime outdoor thermal comfort within urban areas. Tree canopies serve as natural shields, thereby reducing urban surface temperature and resulting in a consistent cooling effect. Additionally, urban greening can also reduce surrounding air temperature through evapotranspiration. However, the tree canopies could create a nighttime warming effect to some extent due to radiative trapping at night (T. Chen et al., 2021). However, this warming effect proves to be marginal compared with the significant cooling effect observed during daytime. The overall benefits of urban tree canopies in improving urban thermal comfort are still substantial.

The characteristics of urban built-up areas, including buildings, roads and other man-made surfaces, affect the budget of both shortwave and longwave radiation within the urban environment. The changed heat capacity and convective heat transfer alter the surface energy budget, deteriorating urban thermal environment (Oke et al., 2017). Specifically, Building Density (BD) emerges as a more influential factor compared to FAR and RSF. The contribution of BD is the most significant around midday, when solar angle is high, even surpassing the contribution of TCF. However, the contribution of urban built-up features is generally less pronounced than those of TCF and WSF at most hours during the day. This suggests that the quantity and area of buildings alone are not the most important factors in affecting pedestrian outdoor thermal comfort. The primary factors contributing to the deterioration of the urban thermal environment are the transformation of the natural surface and the lack of shade provision in these artificial terrains. Drawing from these insights, a holistic strategy for urban planning may involve constructing more high-rise buildings (increasing FAR) to fulfill urban functions, reducing building footprints (decreasing BD), and preserving larger areas of urban forestry (increasing TCF).

## 5. Conclusions

Urban thermal comfort maps at high spatial-temporal resolution are valuable for urban planning and climate research, but most past studies have focused only on block or neighborhood scales due to the complexity of urban morphology and limited sources of weather data. In this study, we proposed an effective framework for city-scale thermal comfort mapping at high spatial-temporal resolution, by coupling the meso-scale numerical weather forecast model (WRF) and the micro-scale thermal environment simulation model (SOLWEIG). Detailed urban spatial information, including building data, tree canopy and ground cover, was taken into account. The framework is applied in the center urban area of Guangzhou, China, to achieve thermal comfort mapping at hourly intervals during a heat wave period. The result can be used for assessing the heat exposure at the city-scale, offering potential for evaluating heat-related risks and evaluating human thermal stress in a complex urban setting.

Results indicate a strong dependence of outdoor thermal comfort on morphological parameters, especially during the daytime when the dependence is even greater than that of local climate factors. The urban outdoor thermal comfort is primarily aggravated by a large proportion of building areas and road surfaces and can be improved by tree canopies and water bodies. When the fraction of tree canopy cover (TCF) reaches 40 %, the UTCI can be reduced by about 2 ºC around noon (9:00 - 13:00). A nighttime warming effect is also observed but is minimal compared to its cooling effect during daytime. The floor area ratio (FAR) is found to be much less influential than building density (BD) and road surface fraction (RSF). Therefore, a suggestion for future urban planning may involve controlling building density and promoting the construction of high-rise buildings, while preserving more areas for urban forestry.

**Acknowledgments**

Research is funded by "Pioneer" and "Leading Goose" R&D Program of Zhejiang (2023C03152) and "Zhejiang University Global Partnership Fund" in collaboration with the Chair of Building Physics at ETH Zurich (100000-11320/209). Research is also supported by The Fundamental Research Funds for the Central Universities (226-2022-00029)

**Appendix A. WRF model configuration**

Essential model configurations are listed in Table A1. The simulation was conducted using the WRF model version 4.4 coupled with the building effect parameterization (BEP) and building energy model (BEM) scheme (Martilli et al., 2002; Salamanca et al., 2010), in a two-way nested domain setup. Four domains at different resolutions were used in this study, and the resolution of each domain is 13.5 km, 4.5 km, 1.5 km, and 0.5 km with 50 vertical levels for each domain. The inner domain (Domain04) is a 121 × 121 grid matrix with a grid cell size of 500 m, while the study area is a 54 × 44 grid matrix within the inner domain. The land cover data used in the outer domain (Domain01-03) was the default MODIS land cover with the 30″ spatial resolution. The local climate zone (LCZ) data (https://www.wudapt.org/) was used for the inner domain (Domain04) simulation to couple with the urban canopy model. The simulation was driven by the NCEP FNL (Final) operational global analysis and forecast data (National Centers for Environmental Prediction, National Weather Service, NOAA, U.S. Department of Commerce, 2015). In order to improve the simulation results, the 4-D data assimilation (FDDA) scheme (Deng & Stauffer, 2006) was activated in coarse domains (Domain01 and 02). Other physical configurations were used according to Zonato et. al. (2020).

**Table A1**

The general configuration of the WRF model.

| Categories | Domian01 | Domian02 | Domian03 | Domian04 |
| --- | --- | --- | --- | --- |
| Resolution | 13.5 km | 4.5 km | 1.5 km | 0.5 km |
| Grids | 100×100 | 121×121 | 118×118 | 121×121 |
| Vertical levels | 50 | 50 | 50 | 50 |
| Land cover data | Default MODIS land cover data | | | LCZ map |
| Urban physics | Noah-MP (multi-physics) Land Surface Model | | | BEP-BEM |
| Data assimilation | True | True | False | False |

**Appendix B. WRF model evaluation**

The WRF-UCM model is validated in terms of 2 m air temperature and 10 m wind speed, using observations from 72 municipal automatic weather stations (AWS). These AWS are operated and maintained by the Guangzhou government in accordance with the national standard "Specifications For Surface Meteorological Observations Standard GB/T 35221-2017".[1] Meteorological observation data used in this study, including air temperature at 2 m height and wind speed 10 m height, were collected from these 72 weather monitoring stations within the study area. Data quality control was performed according to China's quality control standard for surface meteorological observation (Ding et al., 2023). Following the quality control procedures, the raw weather data, recorded at 5-minute intervals, were resampled into hourly mean values. The resampled data achieved an average data integrity rate of 98.9% for 2 m air temperature and 96.0% for 10 m wind speed records. Considering the high data integrity rate, missing data were not filled in or used for any other purpose in this study. Root-mean-square-error (RMSE) at each station was calculated to measure the model performance and compare with other state-of-art studies. In order to evaluate if the model produces overestimated or underestimate predictions, the mean bias between observation and prediction (BIAS) at each hour was also calculated. The mean RMSE at each station and BIAS at each hour are defined in Equation B1 and Equation B2:

$$\text{RMSE} = \sqrt{\frac{\sum_{i=1}^{n}(o_i - p_i)^2}{n}} \tag{B1}$$

$$\text{BIAS} = \frac{\sum_{i=1}^{m}(o_i - p_i)}{m} \tag{B2}$$

where $n$ is the total number of hours in the predicted time series, $m$ is the total number of weather stations used for evaluation, $p_i$ the prediction, and $o_i$ the observation.

The comparison of mean simulation values and observations, as well as the evaluated mean BIAS at each hour, is illustrated in Figure B1a, c. The averaged BIAS for air temperature and wind speed is 0.89 ºC and 0.17 m/s, respectively. The largest bias between observed and predicted air temperature occurs during nighttime, and the model tends to produce underestimated results. As daylight breaks and temperatures rise, the magnitude of BIAS tends diminishes. This discrepancy may be due to the current urban canopy parameterization's inability to account for anthropogenic heat from traffic, which accumulates during evening rush hours, and the underestimation of anthropogenic emission of heat from urban surfaces during the evenings (Berardi et al., 2020). The evaluated RMSE of air temperature and wind speed at each station is shown in Figure B1b, d, and the average RMSE for air temperature and wind speed is 1.65 ºC and 0.78 m/s, respectively. The model demonstrates similar prediction performance at each station during the whole period for both two parameters. For air temperature, most stations have an RMSE ranging from 1.0 to 2.0 ºC, with a standard deviation of 0.44 ºC. Meanwhile, the RMSE of wind speed prediction in most stations is lower than 1.0 m/s, with a standard deviation of 0.4 m/s. Overall, the RMSE score of this WRF-UCM model is comparable to recent studies (Jandaghian & Berardi, 2020; Pappaccogli et al., 2021; Singh et al., 2022; Zonato et al., 2020), which typically range from 1.0 to 2.0 ºC for air temperature and 0.5 to 2.5 m/s for wind speed. Based on these results, we can assume that the simulation is fairly accurate and yields results comparable to state-of-the-art studies.

---

[1] available at https://www.cma.gov.cn/zfxxgk/gknr/flfgbz/bz/202209/t20 220921_5099079.html.

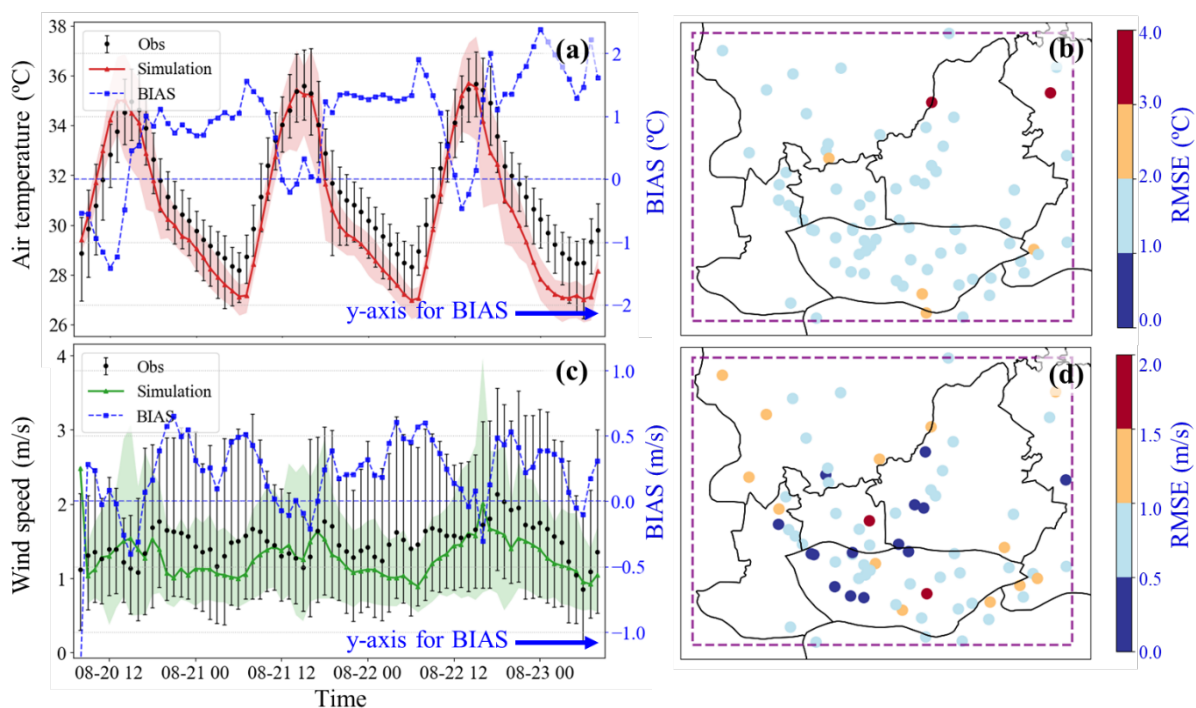

**Figure B1.** Evaluation of WRF simulation. (a, c) Comparison of observed and simulated values, and the mean BIAS at each hour. The error bars and colored area denote 95% confidence intervals of observation and simulation value, respectively. (b, d) the mean RMSE at each weather station.

**Appendix C. Computation of mean radiant temperature**

The mean radiant temperature (MRT) within the study area was simulated using the Solar Long Wave Environmental Irradiance Geometry model (SOLWEIG) (Lindberg et al., 2008), which is implemented in the Urban Multi-scale Environmental Predictor (UMEP) (Lindberg et al., 2018) plugin in the QGIS platform. The SOLWEIG model calculates shortwave and longwave radiation fluxes from six directions individually, with a thorough radiation scheme (Lindberg et al., 2008, 2016). As defined in Equation C1, the mean radiant flux (R) is first calculated as the sum of all fields of shortwave ($K_i$) radiation and longwave ($L_i$) together with the angular ($F_i$) and absorption factors ($\alpha_S$ and $\alpha_L$) of an individual. Then the MRT can be determined using Stefan

Boltzmann's law. The detailed equations to calculate shortwave and longwave radiation are extensively documented in Lindberg et al (Lindberg et al., 2008, 2016; Zheng et al., 2018).

$$R = \alpha_S \sum_{i=1}^{6} K_i F_i + \alpha_L \sum_{i=1}^{6} L_i F_i \tag{C1}$$

The ground cover scheme (Zheng et al., 2018) in the SOLWEIG model was activated to consider the impact of different ground types on the outgoing shortwave and longwave radiation fluxes. The model simulates the effect of land cover on the outgoing shortwave and longwave fluxes by assigning different radiative properties to each land cover type, as listed in Table C1. The outgoing shortwave radiation is further modified by the surface albedo specified by different ground cover types. The longwave radiation fluxes are affected by the surface emissivity and the surface temperature for each surface type. In SOLWEIG, the surface temperature on sun-exposed surfaces on a clear day is assumed to be sinusoidal, and the period of the sinusoidal equation is determined according to the day of the year and the occurrence time of the maximum difference between sunlit surface temperature and air temperature (Lindberg et al., 2008, 2016). Based on the work of Bogren et al (2000), the amplitude and initial morning values are derived from the linear relation presented on the linear relation between maximum solar elevation, $\eta_{max}$, and maximum difference between sunlit surface temperature and air temperature, $\Delta T_{diff\,max}$, as defined in Equation C2. The most frequent occurrences time of $\Delta T_{diff\,max}$ for different ground types are also implemented, which is listed in Table C1.

$$\Delta T_{diff\,max} = k \times \eta_{max} + \Delta T_{initial} \tag{C2}$$

where $k$ is the slope coefficient for different land cover types and $\eta_{max}$ is the maximum sun elevation angle under clear sky conditions. $\Delta T_{initial}$ is the difference between surface temperature and air temperature in early morning before sunrise. Additionally, when the surface is sunlit or shadowed, the surface temperature is assumed to gradually rise or decrease in two sequential hours. For instance, within two hours of shade, the surface temperatures are assumed to gradually return to the temperature of the air.

**Table C1**
Radiative properties of different ground cover types.

| Ground cover type | Albedo | Emissivity | $k$ | $\Delta T_{initial}$ | Time of $\Delta T_{diff\,max}$ (Local time, h) |
|---|---|---|---|---|---|
| Impervious surface | 0.18 | 0.95 | 0.58 | -9.78 | 14 |
| Low vegetation | 0.16 | 0.94 | 0.21 | -3.38 | 15 |

| | | | | |
|---|---|---|---|---|
| Bare soil | 0.25 | 0.94 | 0.33 | -3.01 | 14 |
| Water | 0.05 | 0.98 | 0 | 0 | - |

**Appendix. D Statistical analysis results**

The mean Pearson's correlation coefficients between meteorological, and urban morphological parameters and UTCI values at each hour of the day (Figure D1), and the mean |SHAP| values of Random Forest (RF) regression analysis at each hour of the day. The regression scores, represented by the coefficient of determination ($R^2$), on both train set and test set are close, which means the model is not overfitting. Overall, the model has a better regression score during the daytime than at night.

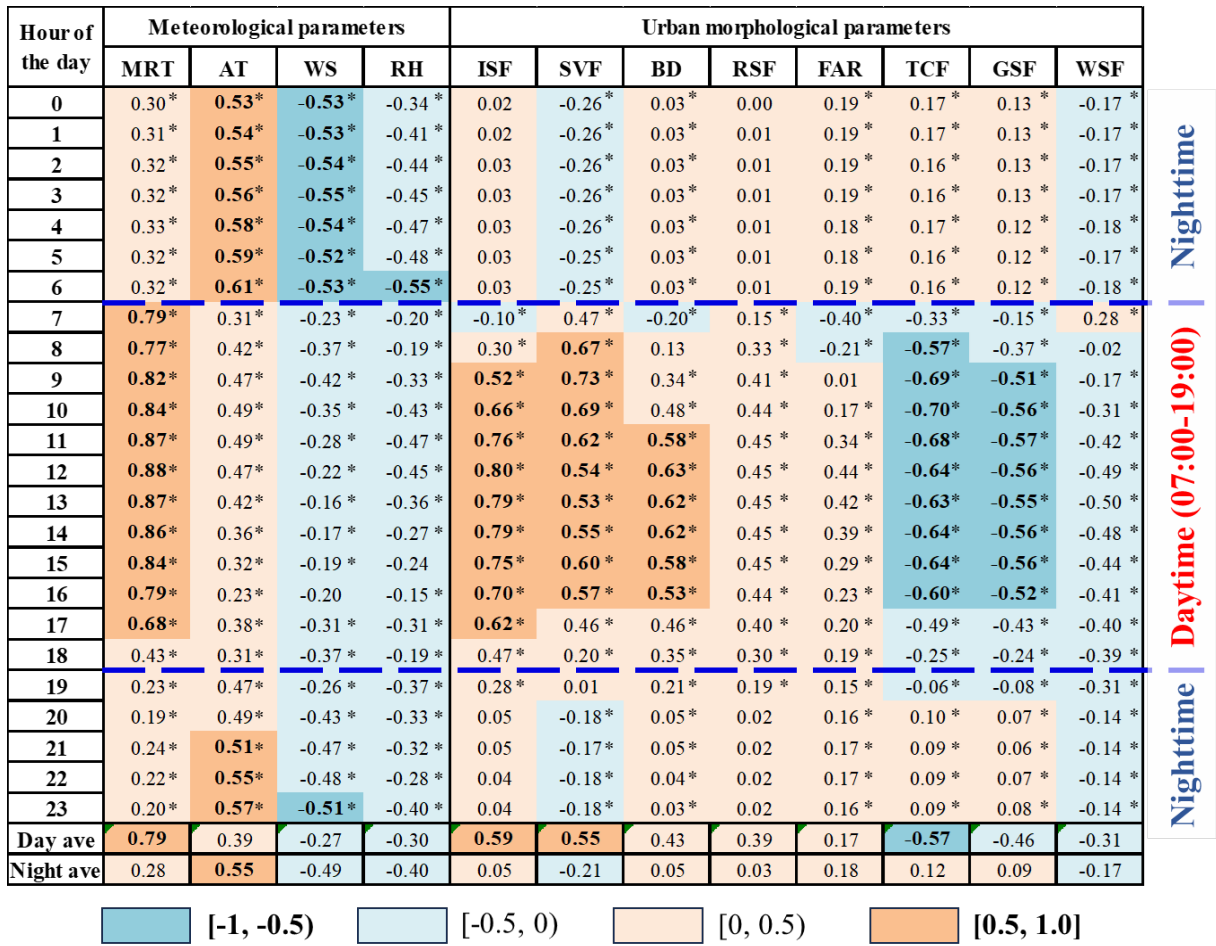

| Hour of the day | Meteorological parameters | | | | Urban morphological parameters | | | | | | | |
|---|---|---|---|---|---|---|---|---|---|---|---|---|
| | MRT | AT | WS | RH | ISF | SVF | BD | RSF | FAR | TCF | GSF | WSF |
| 0 | 0.30* | 0.53* | -0.53* | -0.34* | 0.02 | -0.26* | 0.03* | 0.00 | 0.19* | 0.17* | 0.13* | -0.17* |
| 1 | 0.31* | 0.54* | -0.53* | -0.41* | 0.02 | -0.26* | 0.03* | 0.01 | 0.19* | 0.17* | 0.13* | -0.17* |
| 2 | 0.32* | 0.55* | -0.54* | -0.44* | 0.03 | -0.26* | 0.03* | 0.01 | 0.19* | 0.16* | 0.13* | -0.17* |
| 3 | 0.32* | 0.56* | -0.55* | -0.45* | 0.03 | -0.26* | 0.03* | 0.01 | 0.19* | 0.16* | 0.13* | -0.17* |
| 4 | 0.33* | 0.58* | -0.54* | -0.47* | 0.03 | -0.26* | 0.03* | 0.01 | 0.18* | 0.17* | 0.12* | -0.18* |
| 5 | 0.32* | 0.59* | -0.52* | -0.48* | 0.03 | -0.25* | 0.03* | 0.01 | 0.18* | 0.16* | 0.12* | -0.17* |
| 6 | 0.32* | 0.61* | -0.53* | -0.55* | 0.03 | -0.25* | 0.03* | 0.01 | 0.19* | 0.16* | 0.12* | -0.18* |
| 7 | 0.79* | 0.31* | -0.23* | -0.20* | -0.10* | 0.47* | -0.20* | 0.15* | -0.40* | -0.33* | -0.15* | 0.28* |
| 8 | 0.77* | 0.42* | -0.37* | -0.19* | 0.30* | 0.67* | 0.13 | 0.33* | -0.21* | -0.57* | -0.37* | -0.02 |
| 9 | 0.82* | 0.47* | -0.42* | -0.33* | 0.52* | 0.73* | 0.34* | 0.41* | 0.01 | -0.69* | -0.51* | -0.17* |
| 10 | 0.84* | 0.49* | -0.35* | -0.43* | 0.66* | 0.69* | 0.48* | 0.44* | 0.17* | -0.70* | -0.56* | -0.31* |
| 11 | 0.87* | 0.49* | -0.28* | -0.47* | 0.76* | 0.62* | 0.58* | 0.45* | 0.34* | -0.68* | -0.57* | -0.42* |
| 12 | 0.88* | 0.47* | -0.22* | -0.45* | 0.80* | 0.54* | 0.63* | 0.45* | 0.44* | -0.64* | -0.56* | -0.49* |
| 13 | 0.87* | 0.42* | -0.16* | -0.36* | 0.79* | 0.53* | 0.62* | 0.45* | 0.42* | -0.63* | -0.55* | -0.50* |
| 14 | 0.86* | 0.36* | -0.17* | -0.27* | 0.79* | 0.55* | 0.62* | 0.45* | 0.39* | -0.64* | -0.56* | -0.48* |
| 15 | 0.84* | 0.32* | -0.19* | -0.24 | 0.75* | 0.60* | 0.58* | 0.45* | 0.29* | -0.64* | -0.56* | -0.44* |
| 16 | 0.79* | 0.23* | -0.20 | -0.15* | 0.70* | 0.57* | 0.53* | 0.44* | 0.23* | -0.60* | -0.52* | -0.41* |
| 17 | 0.68* | 0.38* | -0.31* | -0.31* | 0.62* | 0.46* | 0.46* | 0.40* | 0.20* | -0.49* | -0.43* | -0.40* |
| 18 | 0.43* | 0.31* | -0.37* | -0.19* | 0.47* | 0.20* | 0.35* | 0.30* | 0.19* | -0.25* | -0.24* | -0.39* |
| 19 | 0.23* | 0.47* | -0.26* | -0.37* | 0.28* | 0.01 | 0.21* | 0.19* | 0.15* | -0.06* | -0.08* | -0.31* |
| 20 | 0.19* | 0.49* | -0.43* | -0.33* | 0.05 | -0.18* | 0.05* | 0.02 | 0.16* | 0.10* | 0.07* | -0.14* |
| 21 | 0.24* | 0.51* | -0.47* | -0.32* | 0.05 | -0.17* | 0.05* | 0.02 | 0.17* | 0.09* | 0.06* | -0.14* |
| 22 | 0.22* | 0.55* | -0.48* | -0.28* | 0.04 | -0.18* | 0.04* | 0.02 | 0.17* | 0.09* | 0.07* | -0.14* |
| 23 | 0.20* | 0.57* | -0.51* | -0.40* | 0.04 | -0.18* | 0.03* | 0.02 | 0.16* | 0.09* | 0.08* | -0.14* |
| Day ave | 0.79 | 0.39 | -0.27 | -0.30 | 0.59 | 0.55 | 0.43 | 0.39 | 0.17 | -0.57 | -0.46 | -0.31 |
| Night ave | 0.28 | 0.55 | -0.49 | -0.40 | 0.05 | -0.21 | 0.05 | 0.03 | 0.18 | 0.12 | 0.09 | -0.17 |

Legend: [-1, -0.5)  [-0.5, 0)  [0, 0.5)  [0.5, 1.0]

Nighttime | Daytime (07:00-19:00) | Nighttime

**Figure D1**. Mean Pearson's correlation coefficients between meteorological, and urban morphological parameters and UTCI values at each hour of the day. Correlations higher than 0.50 are marked in bold font and at the 0.01 significant level (p-value < 0.01) are marked with *.

| Hour of the day | Meteorological parameters | | Urban morphological parameters | | | | RF R² (train set) | RF R² (test set) | |
|---|---|---|---|---|---|---|---|---|---|
| | AT | WS | BD | FAR | RSF | TCF | | | |
| 0 | 0.26 | 0.14 | 0.03 | 0.05 | 0.01 | 0.17 | **0.56** | 0.45 | Nighttime |
| 1 | 0.25 | 0.14 | 0.03 | 0.03 | 0.02 | 0.17 | **0.56** | 0.44 | |
| 2 | 0.30 | 0.12 | 0.03 | 0.04 | 0.02 | 0.17 | **0.57** | 0.46 | |
| 3 | 0.31 | 0.10 | 0.02 | 0.03 | 0.01 | 0.16 | **0.58** | 0.45 | |
| 4 | 0.32 | 0.06 | 0.02 | 0.05 | 0.01 | 0.16 | **0.56** | **0.53** | |
| 5 | 0.33 | 0.05 | 0.02 | 0.05 | 0.01 | 0.17 | **0.59** | 0.46 | |
| 6 | 0.37 | 0.05 | 0.02 | 0.05 | 0.01 | 0.16 | **0.60** | **0.51** | |
| 7 | 0.29 | 0.05 | 0.04 | 0.31 | 0.02 | 0.38 | **0.64** | **0.52** | Daytime (07:00-19:00) |
| 8 | 0.22 | 0.21 | 0.01 | 0.15 | 0.02 | 0.45 | **0.75** | **0.71** | |
| 9 | 0.23 | 0.13 | 0.02 | 0.12 | 0.02 | **0.77** | **0.83** | **0.78** | |
| 10 | 0.27 | 0.08 | 0.05 | 0.07 | 0.04 | **0.77** | **0.84** | **0.77** | |
| 11 | 0.33 | 0.03 | 0.10 | 0.03 | 0.04 | **0.75** | **0.84** | **0.80** | |
| 12 | 0.30 | 0.02 | 0.48 | 0.05 | 0.10 | **0.55** | **0.84** | **0.81** | |
| 13 | 0.30 | 0.03 | **0.63** | 0.02 | 0.18 | 0.47 | **0.83** | **0.78** | |
| 14 | 0.30 | 0.07 | 0.41 | 0.04 | 0.08 | **0.61** | **0.85** | **0.79** | |
| 15 | 0.24 | 0.11 | 0.23 | 0.03 | 0.06 | **0.62** | **0.82** | **0.77** | |
| 16 | 0.11 | 0.12 | 0.19 | 0.06 | 0.06 | **0.52** | **0.77** | **0.71** | |
| 17 | 0.15 | 0.14 | 0.28 | 0.05 | 0.09 | 0.29 | **0.72** | **0.62** | |
| 18 | 0.19 | 0.19 | 0.13 | 0.04 | 0.04 | 0.10 | **0.61** | 0.50 | |
| 19 | 0.28 | 0.09 | 0.05 | 0.02 | 0.02 | 0.02 | **0.52** | 0.44 | Nighttime |
| 20 | 0.26 | 0.15 | 0.01 | 0.02 | 0.02 | 0.11 | **0.52** | 0.43 | |
| 21 | 0.26 | 0.16 | 0.01 | 0.04 | 0.02 | 0.12 | **0.55** | 0.42 | |
| 22 | 0.28 | 0.13 | 0.02 | 0.03 | 0.02 | 0.11 | **0.56** | 0.43 | |
| 23 | 0.31 | 0.09 | 0.03 | 0.03 | 0.02 | 0.11 | **0.56** | 0.44 | |
| Day ave | 0.25 | 0.09 | 0.22 | 0.08 | 0.06 | **0.56** | **0.79** | **0.73** | |
| Night ave | 0.29 | 0.12 | 0.03 | 0.04 | 0.02 | 0.13 | **0.56** | 0.45 | |

0 — 1.0

**Figure D2.** The mean |SHAP| values of RF. The $R^2$ columns represent the fitness of the RF model on the train set and test set, and values higher than 0.5 are marked in bold blue font.

**Declaration of competing interest**
None.

**References**

Abd Elraouf, R., Elmokadem, A., Megahed, N., Abo Eleinen, O., & Eltarabily, S. (2022). The impact of urban geometry on outdoor thermal comfort in a hot-humid climate. Building and Environment, 225(July), 109632. https://doi.org/10.1016/j.buildenv.2022.109632

ASHRAE Fundamentals Handbook 2001 (SI Edition). (2001). American Society of Heating, Refrigerating, and Air- Conditioning Engineers.

Battista, G., Carnielo, E., & De Lieto Vollaro, R. (2016). Thermal impact of a redeveloped area on localized urban microclimate: A case study in Rome. Energy and

Buildings, 133, 446–454. https://doi.org/10.1016/j.enbuild.2016.10.004

Berardi, U., Jandaghian, Z., & Graham, J. (2020). Effects of greenery enhancements for the resilience to heat waves: A comparison of analysis performed through mesoscale (WRF) and microscale (Envi-met) modeling. *Science of the Total Environment*, *747*, 141300. https://doi.org/10.1016/j.scitotenv.2020.141300

Błazejczyk, K. (2011). Mapping of UTCI in local scale (the case of Warsaw). *Prace i Studia Geograficzne*, *47*(January 2011), 275–283.

Bogren, J., Gustavsson, T., Karlsson, M., Postgård, U., & Centre, E. S. (2000). The impact of screening on road surface temperature. *Meteorlogical Applications*, *7*(2), 97–104. https://doi.org/10.1017/s135048270000150x

Bröde, P., Fiala, D., Blazejczyk, K., Epstein, Y., Holmér, I., Jendritzky, G., et al. (2009). Calculating UTCI equivalent temperature. *In: JW Castellani & TL Endrusick, Eds. Proceedings of the 13th International Conference on Environmental Ergonomics, USARIEM, Natick, MA (5pp. on CD-Rom).*

Campbell I, Sachar S, Meisel J, N. R. (2021). *Beating the heat: A sustainable Cooling Handbook for Cities. UN environment programme*.

Cao, Q., Luan, Q., Liu, Y., & Wang, R. (2021). The effects of 2D and 3D building morphology on urban environments: A multi-scale analysis in the Beijing metropolitan region. *Building and Environment*, *192*(55), 107635. https://doi.org/10.1016/j.buildenv.2021.107635

Chen, L., & Ng, E. (2011). Quantitative urban climate mapping based on a geographical database: A simulation approach using Hong Kong as a case study. *International Journal of Applied Earth Observation and Geoinformation*, *13*(4), 586–594. https://doi.org/10.1016/j.jag.2011.03.003

Chen, T., Pan, H., Lu, M., Hang, J., Lam, C. K. C., Yuan, C., & Pearlmutter, D. (2021). Effects of tree plantings and aspect ratios on pedestrian visual and thermal comfort using scaled outdoor experiments. *Science of the Total Environment*, *801*, 149527. https://doi.org/10.1016/j.scitotenv.2021.149527

Deng, A., & Stauffer, D. R. (2006). On improving 4-km mesoscale model simulations. *Journal of Applied Meteorology and Climatology*, *45*(3), 361–381. https://doi.org/10.1175/JAM2341.1

Ding, X., Fan, Y., Li, Y., & Ge, J. (2022). Domain adaptive deep learning models on urban surface classification (submitted for publication).

Ding, X., Zhao, Y., Fan, Y., Li, Y., & Ge, J. (2023). Machine learning-assisted mapping of city-scale air temperature: Using sparse meteorological data for urban climate modeling and adaptation. *Building and Environment*, *234*, 110211. https://doi.org/10.1016/j.buildenv.2023.110211

Du, J., Sun, C., Xiao, Q., Chen, X., & Liu, J. (2020). Field assessment of winter outdoor 3-D radiant environment and its impact on thermal comfort in a severely cold region. *Science of the Total Environment*, *709*(66), 136175. https://doi.org/10.1016/j.scitotenv.2019.136175

Du, S., Zhang, X., Jin, X., Zhou, X., & Shi, X. (2022). A review of multi-scale modelling, assessment, and improvement methods of the urban thermal and wind environment. *Building and Environment*, *213*(February), 108860. https://doi.org/10.1016/j.buildenv.2022.108860

Eniolu, T., Kong, L., Lau, K. K., Yuan, C., & Ng, E. (2017). A study on the impact of shadow-cast and tree species on in-canyon and neighborhood's thermal comfort. *Building and Environment*, *115*, 1–17. https://doi.org/10.1016/j.buildenv.2017.01.005

Fan, P. Y., Chun, K. P., Mijic, A., Mah, D. N. Y., He, Q., Choi, B., et al. (2022). Spatially-heterogeneous impacts of surface characteristics on urban thermal environment, a case of the Guangdong-Hong Kong-Macau Greater Bay Area. *Urban Climate*, *41*(June 2021), 101034. https://doi.org/10.1016/j.uclim.2021.101034

Fan, Y., Ding, X., Wu, J., Ge, J., & Li, Y. (2021). High spatial-resolution classification of urban surfaces using a deep learning method. *Building and Environment*, *200*(March), 107949. https://doi.org/10.1016/j.buildenv.2021.107949

Follos, F., Linares, C., López-Bueno, J. A., Navas, M. A., Culqui, D., Vellón, J. M., et al. (2021). Evolution of the minimum mortality temperature (1983–2018): Is Spain adapting to heat? *Science of the Total Environment*, *784*, 147233. https://doi.org/10.1016/j.scitotenv.2021.147233

Gao, Y., Zhao, J., & Han, L. (2023). Quantifying the nonlinear relationship between block morphology and the surrounding thermal environment using random forest method. *Sustainable Cities and Society*, *91*(January), 104443. https://doi.org/10.1016/j.scs.2023.104443

Han, L., Zhao, J., Gao, Y., & Gu, Z. (2022). Prediction and evaluation of spatial distributions of ozone and urban heat island using a machine learning modified land use regression method. *Sustainable Cities and Society*, *78*(November 2021), 103643. https://doi.org/10.1016/j.scs.2021.103643

Ho, H. C., Knudby, A., Xu, Y., Hodul, M., & Aminipouri, M. (2016). A comparison of urban heat islands mapped using skin temperature, air temperature, and apparent temperature (Humidex), for the greater Vancouver area. *Science of The Total Environment*, *544*, 929–938. https://doi.org/10.1016/j.scitotenv.2015.12.021

Homer, C., Dewitz, J., Jin, S., Xian, G., Costello, C., Danielson, P., et al. (2020). Conterminous United States land cover change patterns 2001–2016 from the 2016 National Land Cover Database. *ISPRS Journal of Photogrammetry and Remote Sensing*, *162*(March), 184–199. https://doi.org/10.1016/j.isprsjprs.2020.02.019

Jamei, E., Rajagopalan, P., Seyedmahmoudian, M., & Jamei, Y. (2016). Review on the impact of urban geometry and pedestrian level greening on outdoor thermal comfort. *Renewable and Sustainable Energy Reviews*, *54*, 1002–1017. https://doi.org/10.1016/j.rser.2015.10.104

Jandaghian, Z., & Berardi, U. (2020). Comparing urban canopy models for microclimate simulations in Weather Research and Forecasting Models. *Sustainable Cities and Society*, *55*(January), 102025. https://doi.org/10.1016/j.scs.2020.102025

Kántor, N., Gál, C. V., Gulyás, Á., & Unger, J. (2018). The Impact of Façade Orientation and Woody Vegetation on Summertime Heat Stress Patterns in a Central European

Square: Comparison of Radiation Measurements and Simulations. *Advances in Meteorology*, *2018*, 1–15. https://doi.org/10.1155/2018/2650642

Kong, F., Chen, J., Middel, A., Yin, H., Li, M., Sun, T., et al. (2022). Impact of 3-D urban landscape patterns on the outdoor thermal environment: A modelling study with SOLWEIG. *Computers, Environment and Urban Systems*, *94*(August 2021), 101773. https://doi.org/10.1016/j.compenvurbsys.2022.101773

Lam, C. K. C., Lee, H., Yang, S. R., & Park, S. (2021). A review on the significance and perspective of the numerical simulations of outdoor thermal environment. *Sustainable Cities and Society*, *71*(March), 102971. https://doi.org/10.1016/j.scs.2021.102971

Li, J., Niu, J., Mak, C. M., Huang, T., & Xie, Y. (2020). Exploration of applicability of UTCI and thermally comfortable sun and wind conditions outdoors in a subtropical city of Hong Kong. *Sustainable Cities and Society*, *52*(August 2019), 101793. https://doi.org/10.1016/j.scs.2019.101793

Lindberg, F., & Grimmond, C. S. B. (2010). Continuous sky view factor maps from high resolution urban digital elevation models. *Climate Research*, *42*(3), 177–183. https://doi.org/10.3354/cr00882

Lindberg, F., Holmer, B., & Thorsson, S. (2008a). SOLWEIG 1.0 - Modelling spatial variations of 3D radiant fluxes and mean radiant temperature in complex urban settings. *International Journal of Biometeorology*, *52*(7), 697–713. https://doi.org/10.1007/s00484-008-0162-7

Lindberg, F., Holmer, B., & Thorsson, S. (2008b). SOLWEIG 1.0 - Modelling spatial variations of 3D radiant fluxes and mean radiant temperature in complex urban settings. *International Journal of Biometeorology*, *52*(7), 697–713. https://doi.org/10.1007/s00484-008-0162-7

Lindberg, F., Onomura, S., & Grimmond, C. S. B. (2016). Influence of ground surface characteristics on the mean radiant temperature in urban areas. *International Journal of Biometeorology*, *60*(9), 1439–1452. https://doi.org/10.1007/s00484-016-1135-x

Lindberg, F., Grimmond, C. S. B., Gabey, A., Huang, B., Kent, C. W., Sun, T., et al. (2018). Urban Multi-scale Environmental Predictor (UMEP): An integrated tool for city-based climate services. *Environmental Modelling and Software*, *99*, 70–87. https://doi.org/10.1016/j.envsoft.2017.09.020

Lundberg, S. M., & Lee, S.-I. (2017). A Unified Approach to Interpreting Model Predictions. In I. Guyon, U. Von Luxburg, S. Bengio, H. Wallach, R. Fergus, S. Vishwanathan, & R. Garnett (Eds.), *Advances in Neural Information Processing Systems* (Vol. 30). Curran Associates, Inc.

Maffenini, L., Pesaresi, M., Freire, S., Politis, P., & Schiavina, M. (2023). GHS-SDATA R2023A - GHS supporting data. European Commission, Joint Research Centre (JRC) [Dataset]. https://doi.org/10.2905/7520C0F6-A54C-41E7-8F13-1EA3ABFAC320

Martilli, A., Clappier, A., & Rotach, M. W. (2002). An urban surface exchange parameterisation for mesoscale models. *Boundary-Layer Meteorology*, *104*(2), 261–304. https://doi.org/10.1023/A:1016099921195

Masson, V., Lemonsu, A., Hidalgo, J., & Voogt, J. (2020). Urban climates and climate change. *Annual Review of Environment and Resources*, *45*, 411–444. https://doi.org/10.1146/annurev-environ-012320-083623

Mohammad, P., Aghlmand, S., Fadaei, A., Gachkar, S., Gachkar, D., & Karimi, A. (2021). Evaluating the role of the albedo of material and vegetation scenarios along the urban street canyon for improving pedestrian thermal comfort outdoors. *Urban Climate*, *40*(April), 100993. https://doi.org/10.1016/j.uclim.2021.100993

Napoli, C. Di, Pappenberger, F., & Cloke, H. L. (2018). Assessing heat-related health risk in Europe via the Universal Thermal Climate Index ( UTCI ). *International Journal of Biometeorology*, *62*, 1155–1165. https://doi.org/https://doi.org/10.1007/s00484-018-1518-2

Napoli, C. Di, Messeri, A., Novák, M., Rio, J., Wieczorek, J., Morabito, M., et al. (2021). The Universal Thermal Climate Index as an Operational Forecasting Tool of Human Biometeorological Conditions in Europe BT - Applications of the Universal Thermal Climate Index UTCI in Biometeorology: Latest Developments and Case Studies. In E. L. Krüger (Ed.) (pp. 193–208). Cham: Springer International Publishing. https://doi.org/10.1007/978-3-030-76716-7_10

National Centers for Environmental Prediction, National Weather Service, NOAA, U.S. Department of Commerce. (2015). NCEP GDAS/FNL 0.25 Degree Global Tropospheric Analyses and Forecast Grids. Boulder CO: Research Data Archive at the National Center for Atmospheric Research, Computational and Information Systems Laboratory.

Oke, T. R., Mills, G., Christen, A., & Voogt, J. A. (2017). *Urban Climates*. Cambridge: Cambridge University Press. https://doi.org/10.1017/9781139016476

Pappaccogli, G., Giovannini, L., Zardi, D., & Martilli, A. (2021). Assessing the Ability of WRF-BEP + BEM in Reproducing the Wintertime Building Energy Consumption of an Italian Alpine City. *Journal of Geophysical Research: Atmospheres*, *126*(8). https://doi.org/10.1029/2020JD033652

Potchter, O., Cohen, P., Lin, T. P., & Matzarakis, A. (2022). A systematic review advocating a framework and benchmarks for assessing outdoor human thermal perception. *Science of the Total Environment*, *833*(October 2021), 155128. https://doi.org/10.1016/j.scitotenv.2022.155128

Salamanca, F., Krpo, A., Martilli, A., & Clappier, A. (2010). A new building energy model coupled with an urban canopy parameterization for urban climate simulations-part I. formulation, verification, and sensitivity analysis of the model. *Theoretical and Applied Climatology*, *99*(3–4), 331–344. https://doi.org/10.1007/s00704-009-0142-9

Singh, V. K., Mughal, M. O., Martilli, A., Acero, J. A., Ivanchev, J., & Norford, L. K. (2022). Numerical analysis of the impact of anthropogenic emissions on the urban environment of Singapore. *Science of the Total Environment*, *806*, 150534. https://doi.org/10.1016/j.scitotenv.2021.150534

Su, Y., Wang, Y., Wang, C., Zhou, D., Zhou, N., Feng, W., & Ji, H. (2022). Coupling relationships between urban form and performance of outdoor environment at the pedestrian level: A case study of Dalian, China. *Building and Environment*, *213*(October 2021), 108514. https://doi.org/10.1016/j.buildenv.2021.108514

Sun, Y., Hu, T., & Zhang, X. (2018). Substantial Increase in Heat Wave Risks in China in a Future Warmer World. *Earth's Future*, *6*(11), 1528–1538. https://doi.org/10.1029/2018EF000963

Takaku, J., Tadono, T., & Tsutsui, K. (2014). Algorithm development of high resolution global DSM generation by ALOS prism. In *2014 IEEE Geoscience and Remote Sensing Symposium* (pp. 4784–4787). https://doi.org/10.1109/IGARSS.2014.6947564

Wai, K. M., Yuan, C., Lai, A., & Yu, P. K. N. (2020). Relationship between pedestrian-level outdoor thermal comfort and building morphology in a high-density city. *Science of the Total Environment*, *708*, 134516. https://doi.org/10.1016/j.scitotenv.2019.134516

Wang, C., Zhan, W., Liu, Z., Li, J., Li, L., Fu, P., et al. (2020). Satellite-based mapping of the Universal Thermal Climate Index over the Yangtze River Delta urban agglomeration. *Journal of Cleaner Production*, *277*, 123830. https://doi.org/10.1016/j.jclepro.2020.123830

Wang, X., Li, H., & Sodoudi, S. (2022). The effectiveness of cool and green roofs in mitigating urban heat island and improving human thermal comfort. *Building and Environment*, *217*(April), 109082. https://doi.org/10.1016/j.buildenv.2022.109082

Wang, Y., Ni, Z., Hu, M., Chen, S., & Xia, B. (2021). A practical approach of urban green infrastructure planning to mitigate urban overheating: A case study of Guangzhou. *Journal of Cleaner Production*, *287*, 124995. https://doi.org/10.1016/j.jclepro.2020.124995

Witte, R. S., & Witte, J. S. (2017). *Statistics*. John Wiley & Sons.

Wu, Z., Qiao, R., Zhao, S., Liu, X., Gao, S., Liu, Z., et al. (2022). Nonlinear forces in urban thermal environment using Bayesian optimization-based ensemble learning. *Science of the Total Environment*, *838*(March), 156348. https://doi.org/10.1016/j.scitotenv.2022.156348

Xu, X., Wu, Y., Wang, W., Hong, T., & Xu, N. (2019). Performance-driven optimization of urban open space configuration in the cold-winter and hot-summer region of China. *Building Simulation*, *12*(3), 411–424. https://doi.org/10.1007/s12273-019-0510-z

Yang, Z., Chen, Y., Zheng, Z., Huang, Q., & Wu, Z. (2020). Application of building geometry indexes to assess the correlation between buildings and air temperature. *Building and Environment*, *167*, 106477. https://doi.org/10.1016/j.buildenv.2019.106477

Yi, P., Liu, L., Huang, Y., Zhang, M., Liu, H., & Bedra, K. B. (2023). Study on the Coupling Relationship between Thermal Comfort and Urban Center Spatial Morphology in Summer. *Sustainability*, *15*(6), 5084. https://doi.org/10.3390/su15065084

Yu, H., Zhang, T., Fukuda, H., & Ma, X. (2023). The effect of landscape configuration on outdoor thermal environment: A case of urban Plaza in Xi'an, China. *Building and Environment*, *231*(November 2022), 110027. https://doi.org/10.1016/j.buildenv.2023.110027

Yu, Z., Chen, S., & Wong, N. H. (2020). Temporal variation in the impact of urban morphology on outdoor air temperature in the tropics: A campus case study. *Building and Environment*, *181*(July), 107132. https://doi.org/10.1016/j.buildenv.2020.107132

Zhang, J., Li, Z., & Hu, D. (2022). Effects of urban morphology on thermal comfort at the micro-scale. *Sustainable Cities and Society*, *86*(June), 104150. https://doi.org/10.1016/j.scs.2022.104150

Zhao, Y., Sen, S., Susca, T., Iaria, J., Kubilay, A., Gunawarden, K., et al. (2023). Beating the Heat: Solution Sets for Developed Cities.

Zhao, Y., Sen, S., Susca, T., Iaria, J., Kubilay, A., Gunawardena, K., et al. (2023). Beating urban heat: Multimeasure-centric solution sets and a complementary framework for decision-making. *Renewable and Sustainable Energy Reviews*, *186*. https://doi.org/10.1016/j.rser.2023.113668

Zheng, S., Guldmann, J. M., Liu, Z., & Zhao, L. (2018). Influence of trees on the outdoor thermal environment in subtropical areas: An experimental study in Guangzhou, China. *Sustainable Cities and Society*, *42*(July), 482–497. https://doi.org/10.1016/j.scs.2018.07.025

Zipper, S. C., Schatz, J., Singh, A., Kucharik, C. J., Townsend, P. A., & Loheide, S. P. (2016). Urban heat island impacts on plant phenology: Intra-urban variability and response to land cover. *Environmental Research Letters*, *11*(5). https://doi.org/10.1088/1748-9326/11/5/054023

Zonato, A., Martilli, A., Di Sabatino, S., Zardi, D., & Giovannini, L. (2020). Evaluating the performance of a novel WUDAPT averaging technique to define urban morphology with mesoscale models. *Urban Climate*, *31*(May 2019), 100584. https://doi.org/10.1016/j.uclim.2020.100584